\documentclass[aps,prd,twocolumn,showpacs,showkeys,amsmath,amssymb]{revtex4}
\usepackage{graphicx}
\usepackage{bm}
\begin{document}

\title{Heavy Pentaquarks}
\author{P.-Z. Huang}
\author{Y.-R. Liu}
\author{W.-Z. Deng}
\author{X.-L. Chen}
\affiliation{%
Department of Physics, Peking University, BEIJING 100871, CHINA}
\author{Shi-Lin Zhu}
\affiliation{
Department of Physics, Peking University, BEIJING 100871, CHINA\\
The Key Laboratory of Heavy Ion Physics, Ministry of Education,
China}

\date{\today}

\begin{abstract}
We construct the spin-flavor wave functions of the possible heavy
pentaquarks containing an anti-charm or anti-bottom quark using
various clustered quark models. Then we estimate the masses and
magnetic moments of the $J^P={1\over 2}^+$ or ${3\over 2}^+$ heavy
pentaquarks. We emphasize the difference in the predictions of
these models. Future experimental searches at BESIII, CLEOc,
BELLE, and LEP may find these interesting states.
\end{abstract}



\maketitle

\pagenumbering{arabic}

\section{Introduction}\label{sec1}

After LEPS Collaboration \cite{leps} announced the discovery of
the narrow $\Theta^+$ pentaquark state around $1540$ MeV, many
other experimental groups
\cite{diana,clas,saphir,itep,clasnew,hermes,svd} claimed that they
confirmed the existence of this exotic baryon with the minimal
quark content $uudd\bar s$. The third component of its isospin is
$I_z=0$. The $pK^+$ spectrum is featureless
\cite{clas,saphir,clasnew,hermes}. So the $\Theta^+$ pentaquark is
an iso-scalar if it is really a member of the anti-decuplet. At
present, the possibility of this state being a member of another
multiplet is not completely excluded. Hence its total isospin is
probably zero. All the other quantum numbers remain undetermined.
Later, another narrow pentaquark candidate $\Xi_5^{--}$ with
strangeness $S=-2$, baryon number $B=1$, and isospin $I=\frac32$
around $1862$ MeV was observed by NA49 Collaboration \cite{na49}.
However this state has not been confirmed by other groups, and its
existence is not established up to now \cite{doubt}.

These exotic baryons are definitely beyond the conventional quark
model, in which ordinary baryons are composed of three quarks and
mesons are composed of a pair of quark and anti-quark. Although
the simple quark model has been extremely successful in the
classification of hadron states, its foundation has not been
derived from Quantum Chromo-dynamics (QCD) so far. QCD allows a
much richer spectrum than that in quark model. For example,
non-conventional hadrons such as glueballs, hybrid mesons, other
multi-qaurk hadrons are expected in QCD. Convinced that the quark
model can not be the whole story, people have been looking for
these exotic hadrons for decades. None of them was established
without controversy \cite{pdg} until the discovery of the
$\Theta^+$ pentaquark.

Theoretical study of pentaquark states dated back to the early
days of QCD using MIT bag model \cite{strot}. A few years ago,
Diakonov et al. predicted the masses and widths of the
anti-decuplet baryons using the chiral soliton model (CSM) and
suggested several reaction channels to look for them \cite{diak},
which partly motivated the experimental search. But the resulting
masses and widths of the anti-decuplet baryons are very sensitive
to the inputs in this model \cite{diak,mp,ellis}. For example,
either adopting the commonly used value $45$ MeV for the
$\sigma$-term or identifying $N(1710)$ as a member of the
anti-decuplet will lead to a $\Xi_5^{--}$ pentaquark with a mass
of $2070$ MeV, which is $210$ MeV higher than that observed by
NA49 Collaboration \cite{diak,page}. Very recently, Ellis et al.
used the new value $(79\pm 7)$ MeV and $(64\pm 7)$ MeV from two
recent analysis \cite{sigmaterm} for the $\sigma$-term and
obtained a fairly good description of both $\Theta^+$ and
$\Xi^{--}$ masses \cite{ellis}. But the theoretical foundation of
the treatment of the pentaquarks in the chiral soliton model is
challenged with the large $N_c$ formalism by Refs.
\cite{cohen,princeton}.

Since early last year, there appeared many theoretical papers
trying to interpret these exotic states. Among them, Jaffe and
Wilczek's (JW) diquark model is a typical one \cite{jaffe}. In
their model, the $\Theta^+$ pentaquark is composed of a pair of
diquarks and a strange anti-quark. The flavor anti-decuplet is
always accompanied by an octet which is nearly degenerate and will
mix with the decuplet. Shuryak and Zahed's suggested that the
pentaquark mass be lower by replacing one scalar diquark with one
tensor diaquark in JW's model \cite{shuryak}. Karliner and
Lipkin's (KL) proposed a diquark-triquark model which also led to
a flavor anti-decuplet and octet \cite{lipkin}. In all the above
clustered quark model, the resulting angular momentum and parity
$J^P$ of the pentaquark can be either ${1\over 2}^+$ or ${3\over
2}^+$. Dudek and Close first estimated the $J={3\over 2}$
$\Theta^+$ pentaquark mass in JW's and KL's model by considering
the spin-orbital force \cite{close}.

Many models (e.g. \cite{carl}) including all the above clustered
quark models were constructed to ensure the pentaquarks possess
positive parity as in the original chiral soliton model
\cite{diak,mp}. But the $\Theta^+$ pentaquark parity is still a
pending issue. For example, QCD sum rule approach \cite{zhu,qsr}
and lattice QCD simulation favor negative parity \cite{lattice}.
Some other models favor negative parity as well
\cite{zhang,carlson}. Recently, many theoretical papers proposed
interesting ways to determine $\Theta^+$ parity
\cite{yu,hosakanew,oh,hanhart,zhaonew,rekalo,nam,mehen}.

In a recent paper \cite{jaffe2}, Jaffe and Wilczek pointed out
that the decay mode $\Xi_5\to \Xi^\ast + \pi$ observed by NA49
\cite{na49} signals the existence of an octet around $1862$ MeV
together the anti-decuplet since the latter can not decay into a
decuplet and an octet in the $SU(3)_f$ symmetry limit. If further
confirmed, this experiment disfavors a $J^P=\frac12^-$ assignment
for $\Xi_5^{--}$ and poses a serious challenge for the chiral
soliton model since there is no baryon pentaquark octet in the
rotational band in this model. But it may be hard to exclude it
since there always exist excited vibrational octet modes. These
modes can not be calculated rigourously within the chiral soliton
model \cite{ellis}.

Besides its mass, the pentaquark magnetic moment is another very
important quantity encoding the underlying quark structure and
dynamics. Practically speaking, it is essential in the calculation
of the cross-sections of the pentaquark electro- and
photo-production processes. Different models may yield the same
mass, especially when experimental data are available. But they
may yield very different results for the magnetic moment, which
provides a very good way to distinguish various models.

We have employed the light cone QCD sum rule technique to
calculate the $\Theta^+$ pentaquark magnetic moment \cite{huang}.
In Ref. \cite{mm,liww} we have calculated magnetic moments of both
${1\over 2}^+$ and ${3\over 2}^+$ octet and anti-decuplet
pentaquark states in the framework of various clustered quark
models. Several other groups have also discussed the light
pentaquark magnetic moments, especially $\mu_{\Theta^+}$
\cite{zhao,mag,hosaka,bijker}.

In this work we extend the same formalism to discuss the heavy
pentaquarks containing an anti-charm or anti-bottom quark
systematically. In the framework of the above clustered quark
models, we get one $SU(3)_f$ anti-sextet in JW's model, both a
sextet and triplet in KL's models with either $J^P={1\over 2}^+$
or ${3\over 2}^+$. Then we estimate their masses and magnetic
moments. Experimentally, several groups have proposed to hunt
heavy pentaquarks at LEP and B factories \cite{armstrong}, which
calls for theoretical efforts on this topic. There have been some
discussions of heavy pentaquarks in Ref.
\cite{jaffe,lipkin,close,lipkin2}.

As in the case of conventional heavy hadrons, the presence of the
heavy anti-quark makes the treatment of the system simpler. If the
light quarks are really strongly correlated as proposed in the
clustered quark models, the heavy pentaquark system is the ideal
place to study this kind of correlation without the additional
complication due to the extra light anti-quark. Exploration of
these heavy exotic states will deepen our knowledge of strong
interactions.

This paper is organized as follows. In Section \ref{sec1} we
review this rapidly developing field briefly. In the following
sections, we calculate the masses and magnetic moments using
various clustered quark models. Finally, we compare the results
from different models and discuss the experimental searches of
these interesting heavy pentaquarks.

\section{Heavy Pentaquark States In Jaffe and Wilczek's Model}\label{sec2}

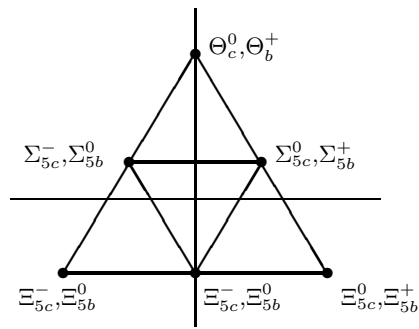
\begin{figure}[h]
\begin{center}
\begin{picture}(200,130)

\thicklines \put(50,22){\line(3,5){50}}
\put(100,105){\line(3,-5){50}} \put(50,22){\line(1,0){100}}
\put(100,22){\line(3,5){25}}\put(100,22){\line(-3,5){25}}\put(75,64){\line(1,0){50}}
\put(50,22){\circle*{4}}\put(100,22){\circle*{4}}\put(150,22){\circle*{4}}
\put(75,64){\circle*{4}}\put(125,64){\circle*{4}}\put(100,105){\circle*{4}}

\thinlines
\put(30,50){\line(1,0){140}}\put(100,2){\line(0,1){120}}

\put(105,105){$\Theta_c^0$,$\Theta_b^+$}
\put(130,64){$\Sigma_{5c}^0$,$\Sigma_{5b}^+$}
\put(35,64){$\Sigma_{5c}^-$,$\Sigma_{5b}^0$}
\put(155,10){$\Xi_{5c}^0$,$\Xi_{5b}^+$}
\put(103,10){$\Xi_{5c}^-$,$\Xi_{5b}^0$}
\put(33,10){$\Xi_{5c}^-$,$\Xi_{5b}^0$}
\end{picture}
\end{center}
\caption{The six members of the $SU(3)$ flavor
anti-sextet.}\label{fig1}
\end{figure}

Jaffe and Wilczek proposed that there exists strong correlation
between the light quark pair when they are in the anti-symmetric
color $({\bar 3}_c)$, flavor $({\bar 3}_f)$, isospin $(I=0)$ and
spin $(J=0)$ configuration \cite{jaffe,jaffe2}. The lighter the
quarks, the stronger the correlation, which helps the light quark
pair form a diquark. For example, the ud diquark behaves like a
scalar with positive parity. Such correlation may arise from the
color spin force from the one gluon exchange or the flavor spin
force induced by the instanton interaction. In order to
accommodate the $\Theta^+$ pentaquark, Jaffe and Wilczek required
the flavor wave function of the diquark pair to be symmetric ${\bf
{\bar 6_f}}$ and their color wave function to be antisymmetric
${\bf 3}_c$. Bose statistics of the scalar diquarks demands an odd
orbital excitation between the two diquarks while there is no
orbital excitation inside each diquark, which ensures that the
resulting pentaquark parity is positive. The heavy anti-quark is a
$SU(3)_f$ flavor singlet. Hence the heavy pentaquarks containing
an anti-charm or anti-bottom form a $SU(3)$ flavor anti-sextet as
shown in Figure \ref{fig1}. The flavor wave functions of all
members of ${\bf {\bar 6_f}}$ are listed in Table \ref{tab1}.

\begin{table}[h]
\begin{center}
\begin{tabular}{c|c|c}\hline
pentaquarks                    &    $(Y,I,I_3)$                              & Flavor wave functions($\bar Q=\bar c$ or $\bar b$) \\
\hline
$\Theta^0_c$,$\Theta^+_b$      &   ($\frac{4}{3}$,0,0)                       &     $[ud]^2 \bar{Q}$      \\
$\Sigma^0_{5c}$,$\Sigma^+_{5b}$& ($\frac{1}{3}$,$\frac{1}{2}$,$\frac{1}{2}$) &     $[ud][us]_+\bar{Q}$   \\
$\Sigma^-_{5c}$,$\Sigma^0_{5b}$& ($\frac{1}{3}$,$\frac{1}{2}$,-$\frac{1}{2}$)&     $[ud][ds]_+\bar{Q}$   \\
$\Xi^0_{5c}$,$\Xi^+_{5b}$      & (-$\frac{2}{3}$,1,1)                        &     $[us]^2\bar{Q}$       \\
$\Xi^-_{5c}$,$\Xi^0_{5b}$      &(-$\frac{2}{3}$,1,)                          &     $[us][ds]_+\bar{Q}$   \\
$\Xi^{--}_{5c}$,$\Xi^-_{5b}$   &(-$\frac{2}{3}$,1,-1)                        &     $[ds]^2\bar{Q}$       \\
 \hline
\end{tabular}
\end{center}
\caption{Flavor wave functions in Jaffe and Wilczek's model
\cite{jaffe}.
$[q_1q_2][q_3q_4]_+=\sqrt{\frac{1}{2}}([q_1q_2][q_3q_4]+[q_3q_4][q_1q_2])$
or $[q_1q_2]^2=[q_1q_2][q_1q_2]$ is the diquark-diquark part.
}\label{tab1}
\end{table}

The magnetic moment of a compound system is the sum of orbital and
spin contributions from all of its constituents. Only the orbital
motion of the scalar diquarks contributes to the pentaquark
magnetic moment. Each diquark's orbital angular momentum reads
\begin{equation}
{\bf l}_1={\bf r}_1 \times {\bf p}_1 =({\bf
R}+\frac{m_2}{m_1+m_2}{\bf r}) \times (\frac{m_1}{m_1+m_2}{\bf
P}+{\bf p})
\end{equation}
where ${\bf R}\equiv(m_1 {\bf r}_1 + m_2 {\bf r}_2)/(m_1 + m_2)$
is the position of the center of mass of the diquark-diquark
system and ${\bf r}\equiv{\bf r}_1 - {\bf r}_2$ is the relative
position between the two diquarks. Since the pentaquark is a bound
state of multi-quarks, $\langle {\bf p_i} \rangle= 0$. So we have
$\langle {\bf P} \rangle =\langle {\bf p_1} \rangle+\langle {\bf
p_2} \rangle= 0$ and $\langle {\bf p} \rangle =\langle (m_2 {\bf
p}_1 - m_1 {\bf p}_2)/(m_1 + m_2)\rangle=0$, from which we get
\begin{eqnarray}
\langle {\bf l}_1 \rangle &=&\frac{m_2}{m_1+m_2} \langle {\bf
r}\times {\bf p} \rangle
+\frac{m_1}{m_1+m_2} \langle {\bf R}\times {\bf P} \rangle \nonumber \\
&=&\frac{m_2}{m_1+m_2} \langle {\bf l} \rangle
+\frac{m_1}{m_1+m_2} \langle {\bf L} \rangle
\end{eqnarray}
where ${\bf l}={\bf r}\times {\bf p}$ is the diquark-diquark's
relative orbital angular momentum.  ${\bf L}={\bf R}\times {\bf
P}$ is the diquark-diquark's orbital angular momentum for the
motion of center of mass. In the center of mass frame $\langle
{\bf L} \rangle=0$. Therefore we get
\begin{equation}
\langle {\bf l}_1 \rangle = \frac{m_2}{m_1+m_2} \langle {\bf l}
\rangle,
\end{equation}
\begin{equation}
\langle {\bf l}_2 \rangle = \frac{m_1}{m_1+m_2} \langle {\bf l}
\rangle.
\end{equation}

For the magnetic moment of the diquark-diquark system we have
\begin{eqnarray}\nonumber
\mu_l \overrightarrow{l} &=& \frac{m_2 \mu_1}{m_1+m_2}
\overrightarrow{l} +\frac{m_1 \mu_2}{m_1+m_2}
\overrightarrow{l}\;, \\
\mu_l &=& \frac{m_2 \mu_1}{m_1+m_2} +\frac{m_1 \mu_2}{m_1+m_2}.
\end{eqnarray}
where $\mu_i\equiv e_i/2m_i$ is the magneton of the $i$-th
diquark. $\overrightarrow{l}=\overrightarrow{1}$ is the angular
moment between diquarks.

Now we can write down the magnetic moment of a $J^P=\frac{1}{2}^+$
pentaquark as follows,
\begin{eqnarray}
\mu&=& \langle 2\mu_{\bar{Q}}\overrightarrow{\frac{1}{2}}
+\mu_l \overrightarrow{l} \rangle (J_z = \frac12) \nonumber\\
   &=& \left( \langle 1 0 \frac12 \frac12 \mid \frac12 \frac12 \rangle^2
- \langle 1 1 \frac12 -\frac12 \mid \frac12 \frac12 \rangle^2
\right) \mu_{\bar{Q}}\nonumber\\
&&+ \langle 1 1 \frac12 -\frac12 \mid \frac12
\frac12 \rangle^2 \mu_l\nonumber\\
&=&-\frac{1}{3}\mu_{\bar{Q}}+\frac{2}{3}\mu_l.
\end{eqnarray}
Similarly, for $J^P=\frac{3}{2}^+$ pentaquarks, we have
\begin{eqnarray}
\mu&=& \langle 2\mu_{\bar{Q}}\overrightarrow{\frac{1}{2}}
+\mu_l \overrightarrow{l} \rangle (J_z = \frac32) \nonumber\\
   &=& \langle 1 1 \frac12 \frac12 \mid \frac32 \frac32 \rangle^2
   \mu_{\bar{Q}} + \langle 1 1 \frac12 \frac12 \mid \frac32 \frac32
\rangle^2 \mu_l \nonumber\\
&=&\mu_{\bar{Q}}+\mu_l
\end{eqnarray}

We use heavy quark masses $m_c=1710$ MeV, $m_b=5050$ MeV from Ref.
\cite{lipkin} and diquark masses $m_{ud}=420$ MeV,
$m_{us}=m_{ds}=600$ MeV from Ref. \cite{jaffe} to compute the
magnetic moments of these $J^P=\frac{1}{2}^+$ and
$J^P=\frac{3}{2}^+$ pentaquarks. The numerical results are
summarized in Table \ref{tab2}.

\begin{table}[h]
\begin{center}
\begin{tabular}{c|c|c|c|c}\hline
&\multicolumn{2}{c}{Q=c}\vline&\multicolumn{2}{c}{Q=b}\\
\cline{2-5}
  ($Y,I,I_3$) &$J^P={1\over 2}^+$&$J^P={3\over 2}^+$&$J^P={1\over 2}^+$&$J^P={3\over 2}^+$\\
\hline
($\frac{4}{3}$,0,0)                        &0.62&0.38&0.48&0.81\\
($\frac{1}{3}$,$\frac{1}{2}$,$\frac{1}{2}$)&0.56&0.29&0.41&0.71\\
($\frac{1}{3}$,$\frac{1}{2}$,-$\frac{1}{2}$)&0.13&-0.36&-0.015&0.071\\
(-$\frac{2}{3}$,1,1)                        &0.47&0.16&0.33&0.58\\
(-$\frac{2}{3}$,1,0)                       &-0.052&-0.63&-0.19&-0.20\\
(-$\frac{2}{3}$,1,-1)                      &-0.57&-1.41&-0.72&-0.98\\
 \hline
\end{tabular}
\end{center}
\caption{Numerical results of the $J^P=\frac{1}{2}^+$ and
$J^P=\frac{3}{2}^+$ heavy pentaquark magnetic moments in Jaffe and
Wilczek's model with $m_{ud}=420$ MeV, $m_{us}=m_{ds}=600$ MeV
from Ref. \cite{jaffe}. }\label{tab2}
\end{table}

Jaffe and Wilczek estimated $\Theta^0_c$ and $\Theta^+_b$ masses
by replacing the $\bar s$ in the $\Theta^+$ with $\bar c$ and
$\bar b$ respectively. The cost of this replacement is roughly the
mass difference between $\Lambda_c$ and $\Lambda$ baryons. The
$[ud]$ diquark in the $\Lambda_c$ and $\Lambda$ experiences nearly
the same environment as in $\Theta^0_c$ and $\Theta^+$
pentaquarks, especially when the $[ud]$ diquark is viewed as a
tightly bound entity. Thus they got
\begin{eqnarray}\nonumber
M(\Theta^0_c)=M(\Theta^+)+[M(\Lambda_c)-M(\Lambda)]\\\nonumber
M(\Theta^+_b)=M(\Theta^+)+[M(\Lambda^+_b)-M(\Lambda)].\nonumber
\end{eqnarray}

Here we extend their formalism to estimate the masses of all the
other members of ${\bf {\bar 6_f}}$. The heavy pentaquarks in the
same isospin multiplet are degenerate.
\begin{eqnarray}\nonumber
M(\Sigma^0_{5c})=M(N^+_s)+[M(\Lambda_c)-M(\Lambda)]\\\nonumber
M(\Sigma^+_{5b})=M(N^+_s)+[M(\Lambda^+_b)-M(\Lambda)].\nonumber
\end{eqnarray}
Here $N^+_s$ has a quark content $|[ud][us]_+ \bar s\rangle$. Its
mass was estimated to be around $1700$ MeV \cite{jaffe}. For the
heavy pentaquark containing two strange quarks, we use
\begin{eqnarray}\nonumber
M(\Xi^0_{5c})=M(\Xi^{--}_5)+[M(\Xi_c)-M(\Sigma)]\\\nonumber
M(\Xi^+_{5b})=M(\Xi^{--}_5)+[M(\Xi_b)-M(\Sigma)].\nonumber
\end{eqnarray}
We use $1860$ MeV for the $\Xi^{--}_5 $ mass from NA49 experiment
\cite{na49}.

Another simple way to estimate $\Xi^0_{5c}$ ($\Xi^+_{5b}$) mass is
\begin{eqnarray}\nonumber
M(\Xi^0_{5c})=M(\Theta^+)+2[m_{us}-m_{ud}]+m_c-m_s\\\nonumber
M(\Xi^+_{5b})=M(\Theta^+)+2[m_{us}-m_{ud}]+m_b-m_s .\nonumber
\end{eqnarray}
With $m_{ud}=420$ MeV, $m_{us}=600$ MeV \cite{jaffe}, $m_c=1710$
MeV and $m_b=5050$ MeV \cite{lipkin}, the $\Xi_{5c}$ and
$\Xi^+_{5b}$ mass from the second estimate is 3110MeV and 5460MeV
respectively, which is only 25MeV below those from the first
estimate. Given that the errors of both estimates are around $100$
MeV, these two approaches yield quite consistent results, which
are collected in Table \ref{mtab1}.

\begin{table}[h]
\begin{center}
\begin{tabular}{c|c|c|c}\hline
$\bar Q=\bar c$               &   Masses  &     $\bar Q=\bar b$         &  Mass    \\
\hline
$\Theta^0_c$                               & $2710$MeV &          $\Theta^+_b$                    & $6050$MeV  \\
$\Sigma^-_{5c}$,$\Sigma^0_{5c}$            & $2870$MeV &    $\Sigma^0_{5b}$,$\Sigma^+_{5b}$       & $6210$MeV  \\
$\Xi^{--}_{5c}$,$\Xi^-_{5c}$,$\Xi^0_{5c}$  & $3135$MeV &  $\Xi^-_{5b}$,$\Xi^0_{5b}$,$\Xi^+_{5b}$  & $6475$MeV  \\
\hline
\end{tabular}
\end{center}
\caption{Masses of all members of ${\bf {\bar 6_f}}$ in Jaffe and
Wilczek's model. }\label{mtab1}
\end{table}

The mass splitting between $J^P=\frac{1}{2}^+$ light pentaquarks
and their $J^P=\frac{3}{2}^+$ partners are estimated to around
tens of MeVs according to Dudek and Close \cite{close}. The mass
splitting between $J^P=\frac{1}{2}^+$ and $J^P=\frac{3}{2}^+$
multiplets scales as the product of the inverse constituent mass.
Hence for heavy pentaquarks, the mass splitting between
$J^P=\frac{1}{2}^+$ and $J^P=\frac{3}{2}^+$ multiplets will be
even smaller, around $10$ MeV.

\section{Heavy Pentaquark States In Karliner and Lipkin's Model}\label{sec4}

In Karliner and Lipkin's (KL) model  \cite{lipkin}, the pentaquark
is divided into two color non-singlet clusters: a scalar diquark
and a triquark. There is a $P$-wave excitation between the two
clusters. The angular momentum barrier between the two clusters
prevents them from rearranging into the usual baryon and meson
system.

The two quarks in the triquark are in the symmetric ${\bf 6_c}$
representation. They couple with the anti-quark to form an
$SU(3)_c$ triplet ${\bf 3}_c$.  The two quarks are in the
anti-symmetric flavor ${\bf \bar{3}_f}$ representation. The
anti-charm or anti-bottom is a $SU(3)_f$ flavor singlet. Hence the
triquark belongs to a flavor anti-triplet. The spin wave function
of the two quarks inside the triquark is symmetric. The spin of
the triquark is one half.

The direct product of the ${\bf \bar 3_f}$ of diquark and the
${\bf \bar 3_f}$ of triquark leads to ${\bf\bar 6_f}$ and ${\bf
3_f}$ pentaquarks. There is one orbital angular momentum $L=1$
between the diquark and the triquark. The resulting $J^P$ of the
pentaquark can be either ${1\over 2}^+$ or ${3\over 2}^+$. We list
the flavor wave functions of these two multiplets in Table
\ref{tab4} and \ref{tab5}.

\begin{table}
\begin{center}
\begin{tabular}{c|c|c} \hline
pentaquarks &($Y,I,I_3$)        &  Flavor wave functions
\\ \hline
$\Theta^0_c$,$\Theta^+_b$      &($\frac{4}{3}$,0,0)                           &$[ud]\{ud\bar{Q}\}$        \\
$\Sigma^0_{5c}$,$\Sigma^+_{5b}$&($\frac{1}{3}$,$\frac{1}{2}$,$\frac{1}{2}$)   &$\sqrt{\frac{1}{2}}([ud]\{us\bar{Q}\}+[us]\{ud\bar{Q}\})$\\
$\Sigma^-_{5c}$,$\Sigma^0_{5b}$&($\frac{1}{3}$,$\frac{1}{2}$,-$\frac{1}{2}$)  &$\sqrt{\frac{1}{2}}([ud]\{ds\bar{Q}\}+[ds]\{ud\bar{Q}\})$\\
$\Xi^0_{5c}$,$\Xi^+_{5b}$      &(-$\frac{2}{3}$,1,1)                          &$[us]\{us\bar{Q}\}$        \\
$\Xi^-_{5c}$,$\Xi^0_{5b}$      &(-$\frac{2}{3}$,1,0)                          &$\sqrt{\frac{1}{2}}([us]\{ds\bar{Q}\}+[ds]\{us\bar{Q}\}$     \\
$\Xi^{--}_{5c}$,$\Xi^-_{5b}$   &(-$\frac{2}{3}$,1,-1)                         &$[ds]\{ds\bar{Q}\}$        \\
\hline
\end{tabular}
\caption{Flavor wave functions of the anti-sextet pentaquarks in
Karliner and Lipkin's model \cite{lipkin}. $Y$, $I$ and $I_3$ are
hypercharge, isospin and the third component of isospin
respectively. $\{q_1q_2\bar{Q}\}\equiv[q_1q_2]\bar{Q}$ is the
triquark's flavor wave function.}\label{tab4}
\end{center}
\end{table}

\begin{table}
\begin{center}
\begin{tabular}{c|c|c} \hline
pentaquarks   &($Y,I,I_3$)        &  Flavor wave functions
\\ \hline
$\Sigma'^0_{5c}$,$\Sigma'^+_{5b}$&($\frac{1}{3}$,$\frac{1}{2}$,$\frac{1}{2}$) &$\sqrt{\frac{1}{2}}([ud]\{us\bar{Q}\}-[us]\{ud\bar{Q}\})$\\
$\Sigma'^-_{5c}$,$\Sigma'^0_{5b}$&($\frac{1}{3}$,$\frac{1}{2}$,-$\frac{1}{2}$)&$\sqrt{\frac{1}{2}}([ud]\{ds\bar{Q}\}-[ds]\{ud\bar{Q}\})$\\
$\Xi'^-_{5c}$,$\Xi'^0_{5b}$      &(-$\frac{2}{3}$,0,0)                        &$\sqrt{\frac{1}{2}}([us]\{ds\bar{Q}\}-[ds]\{us\bar{Q}\}$ \\
\hline
\end{tabular}
\caption{Flavor wave functions of the triplet heavy pentaquarks in
Karliner and Lipkin's model \cite{lipkin}. Notations are the same
as in Table \ref{tab4}.}\label{tab5}
\end{center}
\end{table}

The intrinsic magnetic moment of the triquark is defined as
\begin{equation}
g_{tri} \mu_{tri} \overrightarrow{\frac12} = 2\mu_{q_1}
\overrightarrow{\frac{1}{2}}
+2\mu_{q_2}\overrightarrow{\frac{1}{2}}
+2\mu_{\bar{Q}}\overrightarrow{\frac{1}{2}}.
\end{equation}
From the spin structure of the triquark, we get
\begin{eqnarray}
\frac{1}{2}g_{tri}\mu_{tri}&=&\left(\langle 1 0 \frac12 \frac12
\mid \frac12 \frac12 \rangle^2 -\langle 1 1 \frac12 -\frac12 \mid
\frac12 \frac12 \rangle^2 \right) \mu_{\bar Q} \nonumber
\\ &&+\langle 1 1 \frac12 -\frac12 \mid
\frac12 \frac12 \rangle^2 (\mu_{q_1}+\mu_{q_2})
\end{eqnarray}
For the orbital part we have
\begin{equation}
\mu_l = \frac{m_{tri} \mu_{di} + m_{di} \mu_{tri}}{m_{tri} +
m_{di}},
\end{equation}
where $m_{di}$ is the mass of the diquark, $m_{tri}$ is the mass
of the triquark.

For $J^P=\frac{1}{2}^+$, the magnetic moment of the pentaquark is
\begin{eqnarray}\label{eq15}
\mu &=&\left( \langle 10
\frac{1}{2}\frac{1}{2}|\frac{1}{2}\frac{1}{2}\rangle^2- \langle 11
\frac{1}{2}-\frac{1}{2}|\frac{1}{2}\frac{1}{2} \rangle^2 \right)
\frac{1}{2}g_{tri}\mu_{tri}\nonumber\\
 &&+\langle 1 1
\frac{1}{2}-\frac{1}{2}|\frac{1}{2}\frac{1}{2}\rangle^2 \mu_l \nonumber\\
&=&\frac{2}{3}\mu_l-\frac{1}{3}\times(\frac{1}{2}g_{tri}\mu_{tri}).
\end{eqnarray}
For $J^P=\frac{3}{2}^+$, we have
\begin{eqnarray}
\mu &=& \langle 1 1
\frac{1}{2}\frac{1}{2}|\frac{3}{2}\frac{3}{2}\rangle^2 \mu_l +
\langle 11 \frac{1}{2}\frac{1}{2}|\frac{3}{2}\frac{3}{2}\rangle^2
\frac{1}{2}g_{tri}\mu_{tri}\nonumber\\
&=&\mu_l+\frac{1}{2}g_{tri}\mu_{tri}.
\end{eqnarray}

\begin{table}[h]
\begin{center}
\begin{tabular}{c|c|c|c|c}\hline
pentaquarks  &\multicolumn{2}{c}{Q=c}\vline&\multicolumn{2}{c}{Q=b}\\
\cline{2-5}
   &$J^P={1\over 2}^+$&$J^P={3\over 2}^+$&$J^P={1\over 2}^+$&$J^P={3\over 2}^+$\\
\hline
$\Theta^0_c$,$\Theta^+_b$                                          &-0.031 &1.01  &0.079   &0.96  \\
$\Sigma^0_{5c}$,$\Sigma'^0_{5c}$,$\Sigma^+_{5b}$,$\Sigma'^+_{5b}$  &-0.079 &1.06  &0.030   &1.00  \\
$\Sigma^-_{5c}$,$\Sigma'^-_{5c}$,$\Sigma^0_{5b}$,$\Sigma'^0_{5b}$  &-0.084 &-0.26 &-0.0028 &-0.35 \\
$\Xi^0_{5c}$,$\Xi^+_{5b}$                                          &-0.13  &1.11  &-0.020  &1.05  \\
$\Xi^-_{5c}$,$\Xi'^-_{5c}$,$\Xi^0_{5b}$,$\Xi'^0_{5b}$              &-0.14  &-0.22 &-0.054  &-0.30 \\
$\Xi^{--}_{5c}$,$\Xi^-_{5b}$                                       &-0.15  &-1.54 &-0.089  &-1.66 \\
 \hline
\end{tabular}
\end{center}
\caption{Numerical results of heavy pentaquark magnetic moments in
unit of $\mu_N$ in Karliner and Lipkin's model. }\label{tab6}
\end{table}

\begin{table}[h]
\begin{center}
\begin{tabular}{c|c|c|c}\hline
$\bar Q=\bar c$     &   Mass  &     $\bar Q=\bar b$         &  Mass    \\
\hline
$\Theta^0_c$                     & $2990$MeV &          $\Theta^+_b$                    & $6400$MeV  \\
$\Sigma^-_{5c}(\Sigma'^-_{5c})$, $\Sigma^0_{5c}(\Sigma'^0_{5c})$ &
$3165$MeV & $\Sigma^0_{5b}(\Sigma'^0_{5b})$,
                                               $\Sigma^+_{5b}(\Sigma'^+_{5b})$          & $6570$MeV  \\
$\Xi^{--}_{5c}$,$\Xi^-_{5c}(\Xi'^-_{5c})$,$\Xi^0_{5c}$
                                 & $3340$MeV & $\Xi^-_{5b}$,$\Xi^0_{5b}(\Xi'^0_{5b})$,
                                               $\Xi^+_{5b}$                             & $6740$MeV  \\
\hline
\end{tabular}
\end{center}
\caption{Masses of all members of ${\bf {\bar 6_f}}$ and ${\bf {
3_f}}$ in Karliner and Lipkin's model. }\label{mtab3}
\end{table}

The results are listed in Table \ref{tab6}. We use $m_u=m_d=360$
MeV, $m_s=500$ MeV while $m_c=1710$ MeV, $m_b=5050$ MeV as in Ref.
\cite{lipkin}. The mass of each diquark (triquark) is the sum of
its constituent mass.

Karliner and Lipkin have estimated the masses of $\Theta^+$ and
its heavy flavor analogs \cite{lipkin}. The mass of a pentaquark
comes mainly from the masses of its constituent quarks, the
color-spin hyperfine interaction and the P-wave excitation between
the two clusters. The first two parts can be estimated by
comparing them with a relevant baryon-meson system that have the
same quark contents as this pentaquark. The hyperfine energy
difference between these two systems can be figured out using the
$SU(6)$ color-spin algebra and has been given as
$-\frac{1+\zeta_Q}{12}[M(\Delta)-M(N)]$ by Karliner and Lipkin
\cite{lipkin}, where $\zeta_Q\equiv\frac{m_u}{m_Q}$. The P-wave
excitation energy $\delta E^P$ is approximate to the mass
difference between $D_s(2319)$ and $D^*_s(2112)$. This
approximation is based on the observation that the reduced mass of
the diquark-triquark system is close to that of the $D_s$
\cite{lipkin}.

Following this formalism we estimate the masses of other members
of ${\bf {\bar 6_f}}$ and ${\bf 3_f}$. We have
\begin{eqnarray}\nonumber
 \lefteqn{M(\Sigma^-_{5c})=M(\Sigma^0_{5c})=M(\Sigma'^-_{5c})=M(\Sigma'^0_{5c})}\\\nonumber
 &=\frac{1}{2}[M(N)+M(D_s)+M(\Sigma)+M(D)]
 \\\nonumber
 &-\frac{1+\zeta_c}{12}[M(\Delta)-M(N)]+\delta
 E^{P}\\\nonumber
 \lefteqn{M(\Sigma^0_{5b})=M(\Sigma^+_{5b})=M(\Sigma'^0_{5b})=M(\Sigma'^+_{5b})}\\\nonumber
 &=\frac{1}{2}[M(N)+M(B_s)+M(\Sigma)+M(B)]
 \\\nonumber
 &-\frac{1+\zeta_b}{12}[M(\Delta)-M(N)]+\delta
 E^{P}\nonumber
\end{eqnarray}
and
\begin{eqnarray}\nonumber
 \lefteqn{M(\Xi^{--}_{5c})=M(\Xi^-_{5c})=M(\Xi'^-_{5c})=M(\Xi^0_{5c})}\\\nonumber
 &=M(\Sigma)+M(D_s)]-\frac{1+\zeta_c}{12}[M(\Delta)-M(N)]+\delta E^{P}\\\nonumber
 \lefteqn{M(\Xi^-_{5b})=M(\Xi^0_{5b})=M(\Xi'^0_{5b})=M(\Xi^+_{5b})}\\\nonumber
 &=M(\Sigma)+M(B_s)]-\frac{1+\zeta_b}{12}[M(\Delta)-M(N)]+\delta E^{P}.\nonumber
\end{eqnarray}
The numerical values of the masses of all members of ${\bf {\bar
6_f}}$ and ${\bf 3_f}$ are presented in Table \ref{mtab3}.

\section{Discussions}\label{sec6}

In this paper we have studied the masses and magnetic moments of
heavy pentaquarks in the framework of two clustered quark models.
There is only a $SU(3)$ flavor anti-sextet in Jaffe and Wilczek's
model. In contrast, an additional flavor triplet exists in
Karliner and Lipkin's diquark and triquark model. The masses and
magnetic moments of these triplet pentaquarks are equal to their
anti-sextet partners in KL's model. For the light pentaquarks, all
the above two clustered quark models predicted the existence of an
anti-decuplet and an accompanying octet.

Another dramatic difference lies in the prediction of the magnetic
moments of the $J^P=\frac{1}{2}^+$ heavy sextet pentaquarks.  The
magnetic moments of the $J^P=\frac{1}{2}^+$ heavy sextet is tiny
and much smaller in KL's model than those in JW's model. There is
strong cancellation between the magnetic moment of the triquark
and the orbital magnetic moment in a $J^P=\frac{1}{2}^+$ sextet
when they are combined by Clebsch-Gordan coefficients [c.f.
Eq.(\ref{eq15})]. This cancellation is almost exact for bottom
pentaquarks. For the magnetic moments of the $J^P=\frac{3}{2}^+$
sextet, there is also significant difference between KL's model
and JW's models.

The third difference is that the masses of the pentaquarks in KL's
model are about 300 MeV larger than those in JW's model. The
reason is that Karliner and Lipkin assumed that the diquark
(triquark) mass is simply the sum of its constituents. In
contrast, the diquark mass is made much lower than the sum of the
two quarks through the strong correlation between light quarks
when they are in anti-symmetric configuration in JW's model
\cite{jaffe}.

In JW's models, the masses of heavy pentaquarks are about 100 MeV
below the threshold of the strong decay modes if one believes our
rough estimate. Only iso-spin violating strong decays,
electromagnetic decays and weak decays can occur in this case.
These heavy sextet members should be stable. In the case that
their masses are under-estimated by around 100 MeV in our
calculation, they are still barely above threshold. Severe phase
space suppression will make them very narrow even if they have
strong decays.

On the other hand, the pentaquarks in KL's model lie above
threshold and have strong decay modes. For example, the strong
decay $\Xi_{5b}^+\rightarrow B_s^0\Sigma^+$ and
$\Xi_{5b}^+\rightarrow B^+\Xi^0$ may occur in KL's model. However,
they may still be narrow states as the $\Theta^+$ pentaquark even
if strong decay modes exist.

The presence of additional triplet states, different magnetic
moments, different masses and decay modes will play a role in
distinguishing these two clustered quark models. Future
experiments at CLEOc, BESIII, BELLE and LEP will be able to find
these very interesting states if they really exist. Hopefully the
present study will somehow contribute to the experimental search.

This project was supported by the National Natural Science
Foundation of China under Grant 10375003, Ministry of Education of
China, FANEDD and SRF for ROCS, SEM.


\end{document}